\documentclass[aps,prd,amsmath,amssymb,twocolumn,preprintnumbers,nofootinbib]{revtex4-1}

\usepackage{graphicx}
\usepackage{dcolumn}
\usepackage{bm}
\usepackage[bookmarks, pagebackref=false]{hyperref}
\usepackage[usenames,dvipsnames]{xcolor}
\definecolor{rossoCP3}{cmyk}{0,.88,.77,.40}
\definecolor{blaa}{rgb}{0.2,0.2,0.6}
\hypersetup{colorlinks, 
	bookmarksopen, 
	bookmarksnumbered,
	citecolor=blaa, 		
	linkcolor=rossoCP3,	
	urlcolor=rossoCP3,			
}

\usepackage{amsmath}
\usepackage{slashed}
\usepackage{tikz}
\usepackage{graphicx}
\usepackage{pgfplots}
\usepackage{subfigure}
\usepackage{color}
\usepackage{rotating}
\usepackage{gensymb}
\usepackage{simplewick}
\usepackage{enumerate}
\usetikzlibrary{quotes,angles,snakes}
\usetikzlibrary{trees}
\usetikzlibrary{decorations.pathmorphing}
\usetikzlibrary{decorations.markings}
\usetikzlibrary{arrows,shapes,positioning}
\usetikzlibrary{decorations.markings}
\tikzstyle arrowstyle=[scale=1]
\tikzstyle directed=[postaction={decorate,decoration={markings,
    mark=at position .65 with {\arrow[arrowstyle]{stealth}}}}]
\tikzstyle reverse directed=[postaction={decorate,decoration={markings,
    mark=at position .65 with {\arrowreversed[arrowstyle]{stealth};}}}]

\newcommand{\RNum}[1]{\uppercase\expandafter{\romannumeral #1\relax}}

\newcommand{\beq}{\begin{equation}}
\newcommand{\eeq}{\end{equation}}
\newcommand{\bea}{\begin{eqnarray}}
\newcommand{\eea}{\end{eqnarray}}
\usepackage{ulem}

\newcommand\vRto{\mathrel{\stackrel{\makebox[0pt]{\mbox{\normalfont\tiny \text{$v_R$}}}}{\text{$\longrightarrow$}}}}

\usepackage{xcolor,colortbl}
\definecolor{Blue}{RGB}{140,165,195}
\definecolor{Purple}{RGB}{255,145,145}
\definecolor{bluc}{cmyk}{1,1,0,0.1}
\definecolor{rossoCP3}{cmyk}{0,.88,.77,.40}
\definecolor{rosso}{cmyk}{0,1,1,0.4}
\definecolor{rossos}{cmyk}{0,1,1,0.55}
\definecolor{rossoc}{cmyk}{0,1,1,0.2}
\definecolor{verdes}{cmyk}{0.92,0,0.59,0.4}

\usepackage{color}
\usepackage{ulem}

\begin{document}

\newcommand{\blue}[1]{{\color{blue}#1}}
\newcommand{\red}[1]{{\color{red}#1}}
 
\title{ \LARGE  \color{rossoCP3}  Safe Clockwork }
\author{F.~Sannino$^{\color{rossoCP3}{\heartsuit,\triangle}}$}
\author{J.~Smirnov$^{\color{rossoCP3}{\heartsuit}}$}
\author{Z.W. Wang$^{\color{rossoCP3}{\heartsuit,\diamondsuit}}$}

\affiliation{$^{\color{rossoCP3}{\heartsuit}}$ {\mbox {$\rm{CP}^3$-Origins, University of Southern Denmark, Campusvej 55}}
5230 Odense M, Denmark \\
$^{\color{rossoCP3}{\diamondsuit}}${\mbox {Department of Physics, University of Waterloo, Waterloo, On N2L 3G1, Canada}}\\
$^{\color{rossoCP3}{\triangle}}$ {\mbox{SLAC, National Accelerator Laboratory, Stanford University, Stanford, CA 94025.
}} \\
}

\begin{abstract} 
{In this letter we demonstrate that safe quantum field theories can accommodate the {\it clockwork} mechanism upgrading it to a fundamental theory a l\'a Wilson.}
Additionally the clockwork mechanism naturally sources Yukawa hierarchies for safe completions of the Standard Model (SM). As proof of concept we investigate a safe Pati-Salam (PS) clockwork structure.  \\
[.3cm]
{\footnotesize  \it Preprint: CP$^3$-Origins-2019-004 DNRF90}
\end{abstract}
\maketitle


The current description of fundamental interactions relies on four-dimensional gauge-Yukawa theories which successfully describe the SM of particle interactions. However not all gauge-Yukawa theories are fundamental, i.e. are free from ultraviolet cutoffs. For example, scalar quantum field theories, similar to the one describing the Higgs sector of the SM,  or U(1) hypercharge are best viewed as low energy effective field theories. In fact, if one tries to remove the cutoff by pushing it to arbitrary high energies the resulting consistent quantum field theory corresponds to a non-interacting ({\it trivial}) one. 

Requiring a given extension of the SM to be non-trivial  has proven effective in constraining its interactions and matter content. The PS model of matter field unification \cite{Pati:1974yy} is a time-honoured example in which one can address the hypercharge triviality issue by embedding it in an asymptotically free theory. From a phenomenological standpoint it can be commended because it does not induce fast proton decay, and it can even be extended to provide a stable proton \cite{FileviezPerez:2016laj}. 

So far, asymptotic freedom has been the well traveled route to resolve the triviality problem. An alternative route is that in which the UV theory acquires an interacting fixed point, before gravity sets in, de facto {\it saving} itself from the presence of a cutoff.  This unexplored route was opened when the first safe gauge-Yukawa theory was discovered \cite{Litim:2014uca}. The proof employed rigorous perturbative methods in the Veneziano-Witten limit that requires a large number of fundamental matter fields and colours with the ratio kept fixed.  We also learnt that, when loosing asymptotic freedom for the gauge coupling, scalars are essential to drive asymptotic safety via Yukawa interactions in perturbation theory. 


To achieve a safe theory with a small number of colours we needed to go beyond the state-of-the-art of the large number of matter fields techniques \cite{PalanquesMestre:1983zy,Gracey:1996he}. The first applications of the large $N_f$ limit appeared in  \cite{Mann:2017wzh,Pelaggi:2017abg} where it was first explored whether the SM augmented by a large number of vector-like fermions can have an ultra-violet fixed-point in all couplings.  It was found in \cite{Pelaggi:2017abg} and later on proved in \cite{Antipin:2018zdg} that while the non-abelian gauge couplings, Higgs quartic and Yukawa coupling can exhibit a safe fixed point, the hypercharge remains troublesome. In fact, for abelian  theories  the fermion mass anomalous dimension diverges at the alleged fixed point  \cite{Antipin:2018zdg} suggesting that a safe extension of the SM, like the asymptotically free counterpart, is best obtained by embedding the SM in a non-abelian gauge structure.   

The first non-abelian safe PS and Trinification embeddings were put forward in  \cite{Molinaro:2018kjz, Wang:2018yer}.  However, in the minimal models, only one generation of SM fermions can be modelled, since all the Yukawa couplings are determined by the same UV fixed point value with no resulting hierarchy at low energy.   
 
Safe field theories, in absence of intermediate particle thresholds, are  protected against the emergence of a hierarchy problem  \cite{Pelaggi:2017wzr,Abel:2017ujy,Abel:2017rwl,Abel:2018fls}. Nevertheless, when considering phenomenological extensions of the SM the introduction of vector-like matter at higher energies, required to provide a safe fixed point, induces a certain degree of fine-tuning in the scalar sector of the SM. 

It is therefore interesting to explore whether one can embed extensions of the SM, aimed at addressing the hierarchy problem, into safe quantum field theories. An interesting attempt  was considered in \cite{Cacciapaglia:2018avr} where composite extensions of the SM were taken to be safe rather than asymptotically free.  

Here we argue that the {\it clockwork} mechanism  \cite{Giudice:2016yja} finds home as four-dimensional safe quantum field theory. This is because concrete realizations of the mechanism require the presence of a large number of new vector-like fermions that is a natural prediction of safe quantum field theories. As proof of concept we consider a safe PS structure   \cite{Molinaro:2018kjz} as natural realization of the clockwork idea. Another benefit of this marriage is the use of the clockwork mechanism to generate Yukawa  hierarchies ~\cite{Alonso:2018bcg}.


We will first introduce the model discuss the large number of matter fields renormalisation group, we will then {demonstrate} the presence of safe fixed points and finally offer our conclusions. 

\section{A Safe Model}

We first briefly review the PS embedding of the SM suggested \cite{Molinaro:2018kjz} and then argue that the extra vector-like fermions can naturally play the role of clockwork gears and in the process {\it we kill two birds with one stone}.

\subsection{Pati-Salam Model}

Consider the time-honored PS gauge symmetry group $G_\text{PS}$ \cite{Pati:1974yy}
\begin{equation}
	G_\text{PS} = SU(4)\otimes SU(2)_{L} \otimes SU(2)_{R}\,,
\end{equation}
with gauge couplings $g_4$, $g_L$ and $g_R$, respectively. Here the gauge group $SU(4)\supset SU(3)_C \otimes U(1)_{B-L}$, where $SU(3)_C$ denotes the SM color gauge group.
The SM quark and lepton fields are unified into the $G_{\rm PS}$ irreducible representations
\begin{eqnarray}
\begin{split}\label{fermionsLR}
	\psi_{L i} &= \left(\begin{array}{cccc} u_L  & u_L & u_L & \nu_L\\ d_L  & d_L & d_L & e_L\end{array}\right)_i \sim  (4,2,1)_i \,, \\ 
	\psi_{R i} &= \left(\begin{array}{cccc} u_R  & u_R & u_R & \nu_R\\ d_R  & d_R & d_R & e_R\end{array}\right)_i \sim (4,1,2)_i \,,  
\end{split}
\end{eqnarray}
where $i=1,2,3$ is a flavor index. 
In order to induce the breaking of $G_{\rm PS}$ to the SM gauge group, we introduce a scalar field $\phi_R$ which transforms as the fermion multiplet $\psi_R$, that is $\phi_R\sim (4,1,2)$:
\begin{equation}
	\phi_R \; = \; \left(\begin{array}{cccc}  \phi_R^u & \phi_R^0 \\ \phi_R^d  & \phi_R^-\end{array} \right)\,,
\end{equation}
where the neutral component $\phi_R^0$ takes a non-zero vev, $v_R\equiv \langle\phi_R^0 \rangle$, such that $G_{\rm PS} \vRto SU(3)_C \otimes SU(2)_L \otimes U(1)_Y$.
We also introduce an additional (complex) scalar field $\Phi\sim (1,2,2)$, with
\begin{eqnarray}
	\Phi & = & \left(\begin{array}{cc} \phi_1^0 & \phi_2^+ \\ \phi_1^- & \phi_2^0 \end{array} \right) \equiv \left(\begin{array}{cc} \Phi_1 & \Phi_2 \end{array}\right)\,,
\end{eqnarray}
which is responsible of the breaking of the EW symmetry.

\subsection{Clockwork Extension}

\begin{figure}[h]
\centering
\includegraphics[width=1.05\columnwidth]{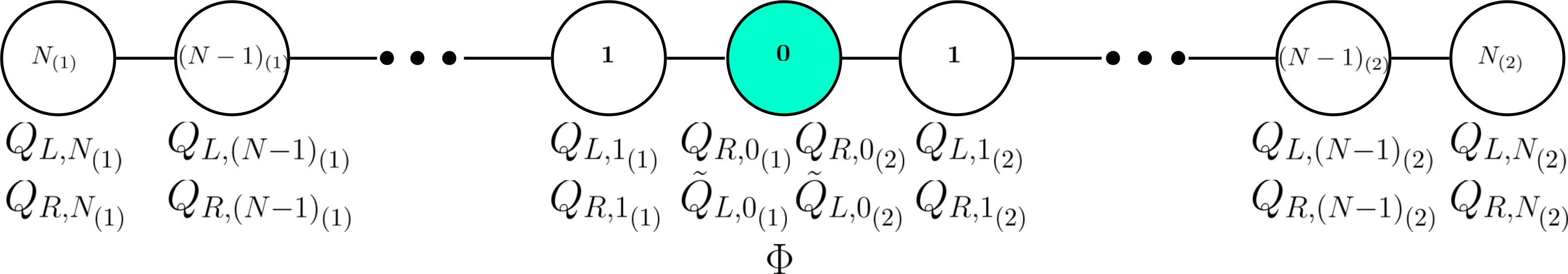}\hspace{0.07\columnwidth}
\caption{The diagram summarises the clockwork chain.}
\label{chain_1}
\end{figure}

To realize the clockwork mechanism, we introduce $N_{\left(i\right)},\,\left(i=1,2\right)$ pair of vector like fermions $\left(Q_{L,1_{(i)}}, Q_{R,1_{(i)}}\right),\,\cdots,\left(Q_{L,N_{\left(i\right)}},Q_{R,N_{\left(i\right)}}\right)$ with one extra chiral fermion $Q_{R,0_{(i)}}$ (i.e.~each generation $(i)$ of the PS fermions is associated with a clockwork chain with $N_{\left(i\right)}$ nodes). Here $\left(i\right)$ denotes the clockwork chain for the first and second generation of the PS (also SM) fermion particles and for any number of $N_{\left(i\right)}$, the chiral fermions $Q_{L,N_{\left(i\right)}}$ and $Q_{R,N_{\left(i\right)}}$ are charged respectively under the fundamental representation $\left(4,2,1\right)$ and $\left(4,1,2\right)$ of PS gauge group $G_{PS}=SU(4)\times SU(2)_L\times SU(2)_R$. In addition, we also introduce $\tilde{Q}_{L,0_{(1)}}$ and $\tilde{Q}_{L,0_{(2)}}$ which will only interact respectively with the zero node fields ${Q}_{R,0_{(1)}}$ and ${Q}_{R,0_{(2)}}$ through the Yukawa contributions i.e.

\begin{equation}
\mathcal{L}_{\rm Yuk}^Q=y_1\bar{\tilde{Q}}_{L,0_{(1)}}\Phi\,Q_{R,0_{(1)}}+y_2\bar{\tilde{Q}}_{L,0_{(2)}}\Phi\,Q_{R,0_{(2)}} \,,\label{LYuk1}
\end{equation}
\\
where $\Phi$ is the scalar bi-doublet in the PS model with charge assignment $\left(1,2,2\right)$. Note that in clockwork type construction of the models, it almost always assumes that the scalar fields are confined to only couple to the fields at one end of the chain where in this work, it is the $0$ node.
The clockwork mechanism is realized by the following clockwork chain interaction (see also Fig.~\ref{chain_1}):
\begin{equation}
\begin{split}
\mathcal{L}_{\rm{clock}}^{Q_R}\,&=-m_{(1)}\sum_{j=1}^{N_{(1)}}\left(\bar{Q}_{L,j_{(1)}}Q_{R,j_{(1)}}-q_{(1)}\bar{Q}_{L,j_{(1)}}Q_{R,{j-1}_{(1)}}\right)\\
&-m_{(2)}\sum_{j=1}^{N_{(2)}}\left(\bar{Q}_{L,j_{(2)}}Q_{R,j_{(2)}}-q_{(2)}\bar{Q}_{L,j_{(2)}}Q_{R,{j-1}_{(2)}}\right)\,,
\end{split}
\end{equation}
where $q_{(i)}>1,\,\left(i=1,2\right)$ is required to realize the clockwork mechanism. For simplicity, in the following we assume $q_{(1)}=q_{(2)}=q$ and $m_{(1)}=m_{(2)}=m$ (i.e.~all the clockwork vectorlike fermions are introduced at one scale). Actually, when turning on the difference between $q$ and $m$, we will have more freedom and bigger parameter space to explore.
After diagonalizing the mass matrix, we obtain $M_{Q_{(i)}}=diag\left(0,M_{1_{(i)}},\cdots,M_{N_{(i)}}\right),\,\left(i=1,2\right)$ where there is always one massless mode $\psi_{R,0_{(i)}},\,\left(i=1,2\right)$.
It is intriguing that the massless modes $\psi_{R,0_{(i)}},\,\left(i=1,2\right)$ overlaps with the fields at the zero node of the chain (see Fig.\,\ref{chain_1}) with a suppression factor i.e.~$\psi_{R,0_{(i)}}=1/q^{N_{(i)}}Q_{R,0_{(i)}}$. Thus, the Yukawa coupling of the $i-$th generations of the PS fermions which originates from the Yukawa 
interaction terms between $\tilde{Q}_{L,0_{(i)}}$ and the massless mode $\psi_{R,0_{(i)}}$ will also be suppressed by $1/q^{N_{(i)}}$ leading to:
\begin{equation}
\begin{split}
\mathcal{L}_{\rm Yuk}^{\rm{eff}}=&\,y_1^{\rm{eff}}\bar{\tilde{Q}}_{L,0_{(1)}}\Phi\,\psi_{R,0_{(1)}}+y_2^{\rm{eff}}\bar{\tilde{Q}}_{L,0_{(2)}}\Phi\,\psi_{R,0_{(2)}}\\
=&\,\frac{y_1}{q^{N_{(1)}}}\bar{\tilde{Q}}_{L,0_{(1)}}\Phi\,\psi_{R,0_{(1)}}+\frac{y_2}{q^{N_{(2)}}}\bar{\tilde{Q}}_{L,0_{(2)}}\Phi\,\psi_{R,0_{(2)}}\,,
\end{split}
\label{LYukeff}
\end{equation}
where it is clear that the effective Yukawa couplings are suppressed and this will be the key to realize the the hierarchies of the Yukawa couplings between different generations of the SM fermions. For reasons that will be presented later in Sec.~\ref{discussion of clockwork} we treat the third generation Yukawa coupling of the PS fermions (i.e. the SM top Yukawa) as the fundamental one while the first and the second generation Yukawa couplings will be generated through the clockwork mechanism.

\section{Large-N$_F$ beta functions}

To the leading $1/N_F$ order, the higher order (ho) contributions to the general RG functions of the gauge couplings  were computed in \cite{Antipin:2018zdg}, while for the simple gauge groups in \cite{Gracey:1996he,Holdom:2010qs} and for the abelian  in \cite{PalanquesMestre:1983zy}. Here we summarize the results. The ho contributions to $d \alpha_i/d\log\mu$ (in the semi-simple case) are:
\begin{equation}
\begin{split}
\beta^{{\rm ho}}_{ i}&=\frac{2A_i\alpha_i}{3}\left(\frac{d(R_i) H_{1_i}(A_i)}{N_{F_i} \,  \prod_{k} d\left(R_\psi^k\right) }+\frac{\sum_{j} \, d(G_j) \,F_{1_j}(A_j)}{N_{F_i} \prod_{k} d\left(R_\psi^k\right) }\right)\,,\\
\alpha_i&\equiv\frac{g_i^2}{\left(4\pi\right)^2}~~\left(i=L,\,R,\,C\right)\,,\label{higher order contribution}
\end{split}
\end{equation}
with the functions $H_{1i}$ and the t'Hooft couplings $A_i$  
\begin{equation}
\begin{split}
A_i&=4\alpha_iT_RN_{F_i}\frac{\prod_{k} d\left(R_\psi^k\right)}{d\left(R_\psi^i\right)}\\ 
H_{1_i}&=\frac{-11}{4}\frac{C_G}{T_R}+\int_0^{A_i/3}I_1(x)I_2(x)dx,\,\\
F_{1_j}&=\int_0^{A_j/3}I_1(x)dx ,\,\\ \label{summation_function_gauge}
\end{split}
\end{equation}
where $I_1(x)$ and $I_2(x)$ are:

\begin{equation}
\begin{split}
I_1(x)&=\frac{\left(1+x\right)\left(2x-1\right)^2\left(2x-3\right)^2\sin\left(\pi x\right)^3}{\left(x-2\right)\pi^3}\\
&\times\left(\Gamma\left(x-1\right)^2\Gamma\left(-2x\right)\right)\\
I_2(x)&=\frac{C_R}{T_R}+\frac{\left(20-43x+32x^2-14x^3+4x^4\right)}{4\left(2x-1\right)\left(2x-3\right)\left(1-x^2\right)}\frac{C_G}{T_R}\,.
\end{split}
\end{equation}
The Dynkin indices are $T_R=1/2~(N_{c_i})$ for the fundamental (adjoint) representation while $d\left(R_\psi^k\right)$ denotes the dimension of the fermion representation.

%
%

The RG functions of the (semi-simple) gauge couplings are:
\begin{equation}
\begin{split}
\beta_{\alpha_{2L}}^{tot}&=\frac{d\alpha_{2L}}{d\log\mu}=\beta_{\alpha_{2L}}^{1loop}+\beta_{\alpha_{2L}}^{\rm{ho}} =-6\alpha_{2L}^2 \\
&+\frac{2A_{2L}\alpha_{2L}}{3}\left(1 + \frac{H_{1_{2L}}\left(A_{2L}\right)}{4\, N_{F}}+\frac{15}{8} \frac{F_{1_4}\left(A_4\right)}{N_F}\right)\\
\beta_{\alpha_{2R}}^{tot}&=\frac{d\alpha_{2R}}{d\log\mu}=\beta_{\alpha_{2R}}^{1loop}+\beta_{\alpha_{2R}}^{\rm{ho}} =-\frac{14}{3}\alpha_{2R}^2 \\
&+\frac{2A_{2R}\alpha_{2R}}{3}\left( 1 + \frac{H_{1_{2R}}\left(A_{2R}\right)}{4\, N_{F}}+\frac{15}{8} \frac{F_{1_4}\left(A_4\right)}{N_F}\right)\\
\beta_{\alpha_{4}}^{tot}&=\frac{d\alpha_{4}}{d\log\mu}=\beta_{\alpha_{4}}^{1loop}+\beta_{\alpha_{4}}^{\rm{ho}} =-18\alpha_{4}^2 \\
+ & \frac{2A_{4}\alpha_{4}}{3}\left( 1 + \frac{H_{1_4}\left(A_{4}\right)}{4\, N_{F}}+ \sum_{i = L/R} \frac{3}{16} \left( \frac{F_{1_{2i}}\left(A_{2i}\right)}{N_F} \right)\right)\,,\\
\end{split}
\label{Gauge couplings RG Bubble}
\end{equation}
The Yukawa beta function reads
\begin{equation}
\begin{split}
\beta_y&=c_1y^3 + y \sum_{\alpha}c_\alpha g^2_\alpha I_y\left(A_\alpha\right),\quad\rm{with}\\\label{eq-simplifiedyukawa}
I_y\left(A_\alpha\right)&=H_\phi\left(0,\tfrac{2}{3}A_\alpha\right)\left(1+A_\alpha\frac{C_2\left(R_\phi^\alpha\right)}{6\left(C_2\left(R_{\chi}^\alpha\right)+C_2\left(R_{\xi}^\alpha\right)\right)}\right)\\
H_\phi(x) &=H_0(x)= \dfrac{(1 - \tfrac{x}{3}) \Gamma(4-x)}{3 \Gamma^2(2 - \tfrac{x}{2}) \Gamma(3 - \tfrac{x}{2}) \Gamma(1 + \tfrac{x}{2})}
\end{split}
\end{equation}
containing information about the resumed fermion bubbles and $c_1,\,c_\alpha$ are the standard 1-loop coefficients for the Yukawa beta function while $C_2(R_\phi^\alpha),\,C_2(R_{\chi}^\alpha),\,C_2(R_{\xi}^\alpha)$ are the Casimir operators of the corresponding scalar and fermion fields. Thus, when $c_1,\,c_\alpha$ are known, the full Yukawa beta function follows.
Similarly, for the quartic coupling we write
\begin{equation}
\begin{split}
\beta_\lambda&=c_1\lambda^2+\lambda \sum_{\alpha}c_\alpha \,g^2_\alpha\,I_{\lambda g^2}\left(A_\alpha\right)+\sum_{\alpha} c'_\alpha \,g_\alpha^4\,I_{g^4} \left(A_\alpha\right) \\
&+\sum_{\alpha < \beta}c_{\alpha\beta}\, g_\alpha^2 g_\beta^2 \,I^{tot}_{g_1^2g_2^2}\left(A_\alpha,\,A_\beta\right)\,,\label{quartic_bubble_beta}
\end{split}
\end{equation}
with $c_1,\,c_\alpha,\,c'_\alpha,\,c_{\alpha\beta}$  the known 1-loop coefficients
for the quartic beta function and the resumed fermion bubbles appear via 
\begin{equation}
\begin{split}
I_{\lambda g^2}\left(A_\alpha\right) &=H_\phi\left(0,\tfrac{2}{3}A_\alpha\right)\\
I_{g^4}\left(A_\alpha\right)&=H_\lambda\left(1,\tfrac{2}{3}A_\alpha\right)+A_\alpha\frac{dH_\lambda\left(1,\tfrac{2}{3}A_\alpha\right)}{dA_\alpha}\\
I_{g_1^2g_2^2}^{tot}\left(A_\alpha,\,A_\beta\right)&=\frac{1}{3}\bigg[I_{g_1^2g_2^2}\left(A_\alpha,\,0\right)+I_{g_1^2g_2^2}\left(0,\,A_\beta\right)\\
&+I_{g_1^2g_2^2}\left(A_\alpha,\,A_\beta\right)\bigg]\\
I_{g_1^2g_2^2}\left(A_\alpha,\,A_\beta\right)&=\frac{1}{A_\alpha-A_\beta}\bigg[A_\alpha H_\lambda\left(1,\tfrac{2}{3}A_\alpha\right)\\
&-A_\beta H_\lambda\left(1,\tfrac{2}{3}A_\beta\right)\bigg],\\
H_\lambda(1,x) &= (1-\tfrac{x}{4}) H_0(x)\\
&=\dfrac{ (1-\tfrac{x}{4})(1 - \tfrac{x}{3}) \Gamma(4-x)}{3 \Gamma^2(2 - \tfrac{x}{2}) \Gamma(3 - \tfrac{x}{2}) \Gamma(1 + \tfrac{x}{2})}\,.
\end{split}
\end{equation} 
We therefore have the  quartic beta function including the bubble diagram contributions when $c_1,\,c_\alpha,\,c_\alpha',\,c_{\alpha\beta}$ are known.

\section{Safe Clockwork Fixed Points}

The clockwork vector-like fermions are charged under $G_{PS}$ with the following charge assignment:
\begin{equation}
N_{F}\left(4,1,2\right)\oplus N_{F}\left(4,2,1\right)\,,\quad N_{F}=N_{(1)}+N_{(2)}\,.
\end{equation}
The physical mass spectrum  of  the vector-like fermions ranges from $m\left(q-1\right)$ to $m\left(q+1\right)$ with $\delta m=2m$ among the states that it is much smaller than the scale hierarchy between the transition scale of the UV fixed point at around $10^{11}\,\rm{GeV}$ and the electroweak scale at $100\,\rm{GeV}$.
It is therefore reasonable to consider, in first approximation, the two generations of vector-like fermions $N_{(1)},\,N_{(2)}$ to appear at the same scale identified with the PS symmetry breaking scale ($\sim1000\,\rm{TeV}$). Thus above the PS symmetry breaking scale, we use directly $N_F$ when analyzing the RG flows. We list the gauge, Yukawa and scalar couplings in Tab.~\ref{couplings}.
\begin{table}[t!]
\centering
  \begin{tabular}{|| l | l | l ||}
    	\hline
Gauge & Yukawa & Scalar \\ \hline
$SU(4):\,g_4$ & $\psi_{L/R}: \,y,\,y_c$ & $\phi_R:\,\lambda_{R1},\,\lambda_{R2}$\\\hline
$SU(2)_L:\,g_L$ & $N_L:\,y_{\nu}$ & portal: $\lambda_{R\Phi_1},\,\lambda_{R\Phi_2},\,\lambda_{R\Phi_3}$\\ \hline
$SU(2)_R:\,g_R$ & $F :\,y_F$ & $\Phi:\,\lambda_1,\,\lambda_2,\,\lambda_3,\,\lambda_4$\\ \hline
\end{tabular}
\caption{\small Gauge, Yukawa and scalar quartic couplings of the PS model.}
\label{couplings}
\end{table}

\begin{table}[t!]
\centering
  \begin{tabular}{|| l | l | l | l | l | l | l | l | l | l | l | l ||}
    	\hline
   	$\lambda_1$ & $\lambda_2$ & $\lambda_3$ & $\lambda_4$ & $\lambda_{R\Phi_1}$ & $\lambda_{R\Phi_2}$ & $\lambda_{R1}$ & $\lambda_{R2}$ & $y$ & $y_c$ & $y_\nu$\\ \hline
   	0.13 & 0.01 & 0.03 & 0.05 & 0.10 & 0.01 & 0.34 & -0.29 & 0.53 & 0.53 & 0.67 \\ \hline
\end{tabular}
\caption{\small This table summarizes the UV fixed point solution for $N_F=13$ 
involving the bubble diagram contributions in the Yukawa and quartic RG beta functions. $y_{F}$ is asymptotically free and thus is zero at the fixed point.
}
\label{shifting UV fixed point_1}
\end{table}

\begin{figure}[h]
\centering
\includegraphics[width=0.9\columnwidth]{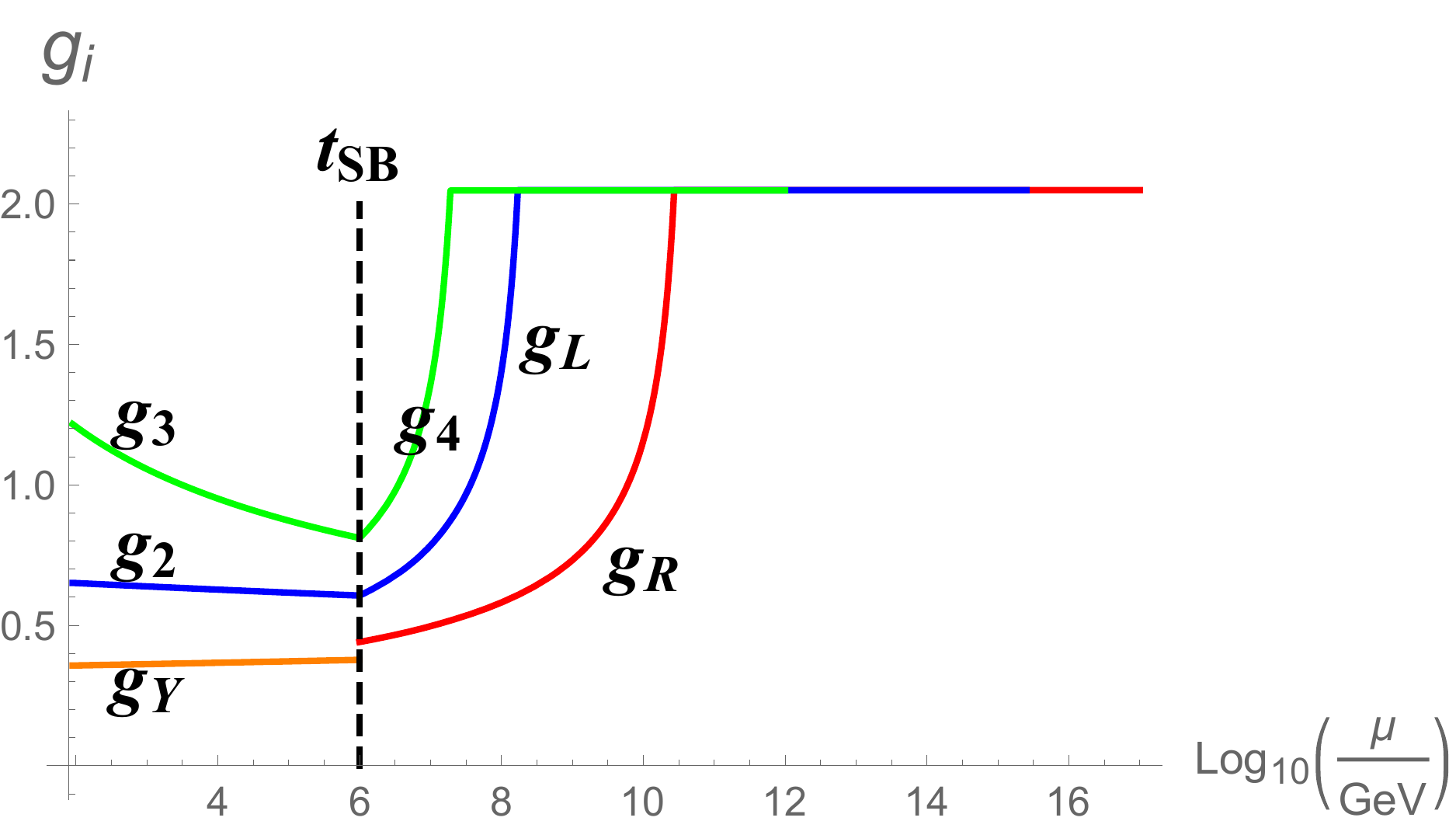}\hspace{0.07\columnwidth}
\caption{The diagram shows the running of the gauge couplings.}
\label{gauge_running}
\end{figure}

For a given value of $N_F$, the gauge couplings at the UV fixed point can be treated as background values (i.e.~constants in the RG functions of other couplings). This is so because the gauge-coupling UV fixed point  depends only on $N_F$ and the group structure. Using the one loop beta functions in \cite{Molinaro:2018kjz} including the large $N_F$ corrections (i.e.~Eq.~\eqref{eq-simplifiedyukawa} and Eq.~\eqref{quartic_bubble_beta}), we solve for $\{\beta_i=0\}$ where $i$ denotes all the Yukawa and scalar couplings in Tab.~\ref{couplings}. A sample UV fixed point solution with $N_F=13$ is shown in Tab.~\ref{shifting UV fixed point_1} where we have selected the solution to satisfy the vacuum stability conditions. In Fig.~\ref{gauge_running}, we show the RG running of the three gauge couplings that achieve a fixed point after $10^{11}\,\rm{GeV}$. It is pleasing that all gauge couplings assume the same value in the UV due to nature and structure of the fixed point.  
To match onto the SM, we consider the RG flows below the PS symmetry breaking scale. 
  After PS symmetry breaking, the scalar bi-doublet should match the conventional two Higgs doublet model which is defined by the Lagrangian:
\begin{equation}
\begin{split}
V_H&=m_{11}^2\Phi_1^\dagger\Phi_1+m_{22}^2\Phi_2^\dagger\Phi_2-\left(m_{12}^2\Phi_1^\dagger\Phi_2+\rm{H.c.}\right)\\
&+\bar{\lambda}_1\left(\Phi_1^\dagger\Phi_1\right)^2+\bar{\lambda}_2\left(\Phi_2^\dagger\Phi_2\right)^2+\bar{\lambda}_3\left(\Phi_1^\dagger\Phi_1\right)\left(\Phi_2^\dagger\Phi_2\right)\\
&+\bar{\lambda}_4\left(\Phi_1^\dagger\Phi_2\right)\left(\Phi_2^\dagger\Phi_1\right)+\bigg[\frac{1}{2}\bar{\lambda}_5\left(\Phi_1^\dagger\Phi_2\right)^2\\
&+\bar{\lambda}_6\left(\Phi_1^\dagger\Phi_1\right)\left(\Phi_1^\dagger\Phi_2\right)+\bar{\lambda}_7\left(\Phi_2^\dagger\Phi_2\right)\left(\Phi_1^\dagger\Phi_2\right)+\rm{H.c.}\bigg]\,,\label{two Higgs doublet}
\end{split}
\end{equation}
and the matching conditions with the scalar couplings in Tab.~\ref{couplings} are (see \cite{Molinaro:2018kjz}):
\begin{equation}
\begin{split}
&\bar{\lambda}_1=\lambda_1,\quad\bar{\lambda}_2=\lambda_1,\quad\bar{\lambda}_3=2\lambda_1,\quad\bar{\lambda}_4=4\left(-2\lambda_2+\lambda_4\right)\\
&\bar{\lambda}_5=4\lambda_2,\quad\bar{\lambda}_6=-\lambda_3,\quad\bar{\lambda}_7=\lambda_3\label{quartic_matching}\,.
\end{split}
\end{equation}
In our system, when $N_F$ is given and the PS symmetry breaking scale is chosen, by using the RG running from the UV fixed point, we obtain the coupling values at the symmetry breaking scale. We can venture below the symmetry breaking scale by using the two Higgs doublet beta functions given in \cite{Branco:2011iw}.
Implementing the matching conditions Eq.~\eqref{quartic_matching}, we treat the coupling values obtained at the PS symmetry breaking scale as the new initial conditions.
We chose the UV fixed point solution shown in Tab.~\ref{shifting UV fixed point_1} as a starting point and PS symmetry breaking scale at $1000\,\rm{TeV}$ to find:
\begin{equation}
\begin{split}
&\bar{\lambda}_1=0.199,~\bar{\lambda}_2=0.199,~\bar{\lambda}_3=0.145,~\bar{\lambda}_4=0.050,\\
&\bar{\lambda}_5=0.034,~y_{\rm{top}}=0.933\,,\label{couplingvalue_EW}
\end{split}
\end{equation}
where the couplings are all defined at the electroweak scale, and with a phenomenologically acceptable top Yukawa coupling. The neutral Higgses mass matrix reads 
\begin{equation}
\left[
\begin{array}{cc}
\frac{m_{12}^2 v_2}{v_1}+2 \bar{\lambda}_1 v_1^2 & -m_{12}^2+\bar{\lambda}_{345}\,v_1 v_2 \\
-m_{12}^2+\bar{\lambda}_{345}\,v_1 v_2 & \frac{m_{12}^2 v_2}{v_1}+2 \bar{\lambda} _2 v_2^2 \\
\end{array}
\right]\,,\label{Mass_two_doublet}
\end{equation}
where $\bar{\lambda}_{345}\equiv \bar{\lambda} _3+\bar{\lambda}_4+\bar{\lambda} _5$.
Eq.~\eqref{Mass_two_doublet} provides a light Higgs mass at $126.6\,\rm{GeV}$ and a heavier Higgs mass at $274.1\,\rm{GeV}$ when setting $M_{12}>100\,\rm{GeV}$. Note that the scalar mass predictions are $m_{12}$ dependent. When setting $m_{12}=0$, the light Higgs will be massless while the heavy Higgs will be around $125\,\rm{GeV}$. However when slightly increasing the $m_{12}$ parameter, the light Higgs will increase correspondingly until $m_{12}\sim100\,\rm{GeV}$. After that the light Higgs mass freezes at around $125\,\rm{GeV}$ while the heavy Higgs mass increases with $m_{12}$.

Summarising for $N_F=13$ we can match both the Higgs mass and the top Yukawa coupling at the electroweak scale. We  searched the full parameter space in the range of $N_F\in\left(10,\,200\right)$ and $N_F=13,\,14$ and the UV fixed point solutions in Tab.~\ref{shifting UV fixed point_1} agree best with the low energy data. We note that  $y_F$ is  asymptotically free for all  viable solutions. We have therefore provided a safe clockwork completion of the SM.

\section{Light generations and conclusions}\label{discussion of clockwork}

{The mass hierarchies among the SM fermion generations are controlled by the \it{clockwork} parameter $q$.}
The relations among  $q^{N_{(1)}}$, $q^{N_{(2)}}$ and the light quark masses are  \begin{equation}
q^{N_{(1)}}=\frac{m_{top}}{m_{u}},~q^{N_{(2)}}=\frac{m_{top}}{m_{c}},~N_{(1)}+N_{(2)}=13\,,\label{clockwork solution}
\end{equation}
where $m_{top}=173\,\rm{GeV}$, $m_{c}=1.29\,\rm{GeV}$ and $m_{u}=2.3\,\rm{MeV}$. By solving Eq.~\eqref{clockwork solution}, we find 
\begin{equation}
N_{(1)}=9,\quad N_{(2)}=4,\quad q=3.46\,.
\end{equation}
A fair point is whether we have enough flavours to argue for the robustness of the large $N_f$ expansion. Following \cite{Antipin:2018zdg}  a naive estimate  is given by 
\begin{equation}
\begin{split}
SU(4): 2N_F\prod_{k} d\left(R_\psi^k\right)/d\left(R_\psi^i\right)&=52>10N_c=40\\
SU(2)_L: N_F\prod_{k} d\left(R_\psi^k\right)/d\left(R_\psi^i\right)&=52>10N_c=20\\
SU(2)_R: N_F\prod_{k} d\left(R_\psi^k\right)/d\left(R_\psi^i\right)&=52>10N_c=20\,,\\
\end{split}
\end{equation}
where the extra factor of 2 for $SU(4)$ comes from both left and right handed species charged under SU(4). We therefore have that $N_F=13$ satisfies the estimated lower bound of the conformal window. 
%

We note that the mass splitting within the PS multiplets needs to be induced by additional operators in order to match the SM values. A possibility includes the addition of $10-\rm{dim}$ multiplets under the PS symmetry group. However,  a detailed study including these operators goes beyond the scope of the present work. 

We have shown that it is mutual beneficial to embed the clockwork mechanism into safe quantum field theories, since the clockwork offers natural ways to generate the observed Yukawa hierarchies while safe field theories naturally predict a large number of vector-like fields for the clockwork to be operative.

\begin{acknowledgments}
F.S. acknowledges discussions with Jogesh C. Pati. The work is partially supported by the Danish National Research Foundation under the grant DNRF:90. 
\end{acknowledgments}
 


\begin{thebibliography}{99}
\bibitem{Pati:1974yy} 
  J.~C.~Pati and A.~Salam,
  Phys.\ Rev.\ D {\bf 10}, 275 (1974)
  Erratum: [Phys.\ Rev.\ D {\bf 11}, 703 (1975)].


\bibitem{FileviezPerez:2016laj} 
  P.~Fileviez Perez and S.~Ohmer,
  Phys.\ Lett.\ B {\bf 768}, 86 (2017)


\bibitem{Litim:2014uca} 
  D.~F.~Litim and F.~Sannino,
  JHEP {\bf 1412}, 178 (2014)


\bibitem{PalanquesMestre:1983zy} 
  A.~Palanques-Mestre and P.~Pascual,
  Commun.\ Math.\ Phys.\  {\bf 95}, 277 (1984).


\bibitem{Gracey:1996he} 
  J.~A.~Gracey,
  Phys.\ Lett.\ B {\bf 373}, 178 (1996)


\bibitem{Mann:2017wzh} 
  R.~Mann, J.~Meffe, F.~Sannino, T.~Steele, Z.~W.~Wang and C.~Zhang,
  Phys.\ Rev.\ Lett.\  {\bf 119}, no. 26, 261802 (2017)


\bibitem{Pelaggi:2017abg} 
  G.~M.~Pelaggi, A.~D.~Plascencia, A.~Salvio, F.~Sannino, J.~Smirnov and A.~Strumia,
  Phys.\ Rev.\ D {\bf 97}, no. 9, 095013 (2018)


\bibitem{Antipin:2018zdg} 
  O.~Antipin, N.~A.~Dondi, F.~Sannino, A.~E.~Thomsen and Z.~W.~Wang,
  Phys.\ Rev.\ D {\bf 98}, no. 1, 016003 (2018)


\bibitem{Molinaro:2018kjz} 
  E.~Molinaro, F.~Sannino and Z.~W.~Wang,
  Phys.\ Rev.\ D {\bf 98}, no. 11, 115007 (2018)


\bibitem{Wang:2018yer} 
  Z.~W.~Wang, A.~Al Balushi, R.~Mann and H.~M.~Jiang,
  ``Safe Trinification,''
  arXiv:1812.11085 [hep-ph].


\bibitem{Pelaggi:2017wzr} 
  G.~M.~Pelaggi, F.~Sannino, A.~Strumia and E.~Vigiani,
  Front.\ in Phys.\  {\bf 5}, 49 (2017)


\bibitem{Abel:2017ujy} 
  S.~Abel and F.~Sannino,
  Phys.\ Rev.\ D {\bf 96}, no. 5, 056028 (2017)


\bibitem{Abel:2017rwl} 
  S.~Abel and F.~Sannino,
  Phys.\ Rev.\ D {\bf 96}, no. 5, 055021 (2017)


\bibitem{Abel:2018fls} 
  S.~Abel, E.~M\o lgaard and F.~Sannino,
  ``A complete asymptotically safe embedding of the Standard Model,''
  arXiv:1812.04856 [hep-ph].


\bibitem{Cacciapaglia:2018avr} 
  G.~Cacciapaglia, S.~Vatani, T.~Ma and Y.~Wu,
  ``Towards a fundamental safe theory of composite Higgs and Dark Matter,''
  arXiv:1812.04005 [hep-ph].


\bibitem{Giudice:2016yja} 
  G.~F.~Giudice and M.~McCullough,
  JHEP {\bf 1702}, 036 (2017)


\bibitem{Alonso:2018bcg} 
  R.~Alonso, A.~Carmona, B.~M.~Dillon, J.~F.~Kamenik, J.~Martin Camalich and J.~Zupan,
  JHEP {\bf 1810}, 099 (2018)


\bibitem{Holdom:2010qs} 
  B.~Holdom,
  Phys.\ Lett.\ B {\bf 694}, 74 (2011)


\bibitem{Branco:2011iw} 
  G.~C.~Branco, P.~M.~Ferreira, L.~Lavoura, M.~N.~Rebelo, M.~Sher and J.~P.~Silva,
  Phys.\ Rept.\  {\bf 516}, 1 (2012)









  \end{thebibliography}
\end{document}